\documentclass[aps,pre,amssymb,amsmath,twocolumn]{revtex4}
\usepackage{graphicx}

\begin{document}
\title{Machine Learning Domain Adaptation in Spin Models with Continuous Phase Transitions}
\author{Vladislav Chertenkov$^{1,2}$} \email{satankow@yandex.ru}
\author{Lev Shchur$^{1, 2}$} \email{Lev.Shchur@gmail.com}
\affiliation{$^1$ Landau Institute for Theoretical Physics, 142432 Chernogolovka, Russia}
\affiliation{$^2$ HSE University, 101000 Moscow, Russia}

\begin{abstract}
The main question raised in the  article  is whether a neural network trained on a spin lattice model in one universality class   can be used to test a model in another universality class. The quantities of interest are the critical phase transition temperature and the correlation length exponent. In other words, the question of transfer learning is how ``universal'' the trained network is and under what conditions.
For this purpose, we applied a supervised learning procedure to three two-dimensional models for which critical properties are precisely known: the Ising model, the four-state Potts model, and the Baxter-Wu model. We consider two datasets: one with spins configurations and one with binding energy configurations. We find that estimates of the critical temperature agree well with the known results for both datasets, but not with the results of cross-testing using the energy datasets of the two models: the four-state Potts model and the Ising model.  Estimates of the critical length exponent are less regular, and appear to be more accurate for energy datasets. A good example is the cross-testing using the energy dataset between Ising model and Baxter-Wu model in both training and testing directions. 
\end{abstract}

\maketitle

\section{Introduction} 

 Machine learning~\cite{Carleo-2019} is a promising approach with possible applications in statistical physics. It has already been established that supervised learning with classification of the spin distribution  into ferromagnetic (FM) or paramagnetic (PM) phases can yield estimates of the critical temperature~\cite{Carrasquilla-2017}  and correlation length exponent~\cite{Chertenkov-2023} when training and testing the same model. A natural question is: can we train a neural network (NN) with a model in one universality class and use it to estimate the critical temperature and correlation length exponent of a model of another universality class? With the right choice of the data set, we answered this question in the affirmative. 

We present an example where a network trained to classify ordered and disordered phases of a spin lattice model with a known phase transition temperature can be applied to predict whether the tested configurations of another spin model belong to an ordered or disordered phase. Using neural network predictions and finite-dimensional analysis, we estimate the phase transition temperature and the critical correlation length exponent of the other spin model. No information about the critical behaviour of the other model is used in this example. Thus, this example shows that it is fundamentally possible to gain new knowledge by investigating a model with unknown properties. Our approach does not claim to be universal or to solve the problem of transfer of knowledge, it only demonstrates the possibility of such an investigation and gives hope that such investigations are possible in principle.

We use a recently developed finite-size scaling approach~\cite{Chertenkov-2023} to analyze  {\em variation} $V(T)$ of the NN output function $P(T)$ - it has been shown that extracting the correlation length and critical temperature from $V(T)$ is very robust both when applied to different physical models and when using different neural network (NN) architectures,  as long as training and testing are performed on the same model.

To the paper, we show that the problem of training/testing spin distributions involves two aspects. Firstly, there are the physical properties of ordered states, which lie in the degeneracy of the ground state. For the Ising model, the degeneracy is two, and the system below $T_c$ can equally have most spins in the +1 or -1 state. In the case of the Baxter-Wu model, four combinations of spins on triangular plaques give the same ground state energy, and this is a fourfold degeneracy.   In the case of the four-state Potts model, ground state is also fourfold degenerate. The ground state degeneracy separates systems with the same dimensionality and the same number of order parameter components into different universality classes~\cite{Privman-1991}. In practice, this means a different set of values of the critical exponents. It should be noted that the critical temperature is not universal and, in particular, differs for four-state Potts model and Baxter-Wu model, which belong to the same universality class. Interestingly, in the usual view of Hamiltonians, the critical temperature is the same for the Ising model and the Baxter-Wu model, but they belong to different universality class.

Second, systems in the same universality class can differ in the symmetry of the interaction between spins, which is reflected in the peculiarities of the local configurations of spins. For example, to the universality class of the 4-component Potts model~\cite{Potts} belong the Baxter-Wu model~\cite{Baxter-1974}, special cases of the Ashkin-Teller model~\cite{Ashkin-1943}, and the Turban model~\cite{Turban-1982}. All four models have the same set of critical exponents and indistinguishable properties of thermodynamic functions at temperatures in the neighborhood of the critical temperature~\cite{Chertenkov-2021}. From the results of applying machine learning to second-order phase transitions~\cite{Carrasquilla-2017,Chertenkov-2023}, we can conclude that the neural network somehow senses correlations in the system, and it is not clear how NNs will respond to local features of the models. For example, it has been observed that local variations can affect the estimation of the critical behaviour of spin models in the universality class of the two-dimensional Ising model. It was verified that a neural network trained on an isotropic Ising model, when testing Ising models with diagonal~\cite{Onsager-1944} and orthogonal anisotropy~\cite{Houtappel-1950}, correctly predicts the critical behaviour only for not large  deviations from the isotropic case~\cite{DDS-1,DDS-2}.   The listed models in the universality class of the 4-component Potts model have significant differences in local properties, including anisotropic character, and, in addition, their study is complicated by the presence of multiplicative logarithmic corrections that significantly complicate numerical analysis.  An exception is the Baxter-Wu model, in which logarithmic corrections are absent due to the unknown symmetry properties of the model~\cite{Baxter-1974}. Therefore, the pairing of the Ising model and the Baxter-Wu model chosen in the paper is a natural step towards the study of the cross-learning problem in spin lattice models with emphasis on cross-learning between universality classes. 

Our problem is the part of the very broad problem of the transfer learning~\cite{TransferML}, and generally it is connected with the problem of domain adaptation, which arises when there are several related domains containing a shift associated with differences in feature space distributions. 
Snapshots of spin distributions for different models at temperatures below the critical temperature tend to be very different from each other, and NN cannot correctly predict the probability that specific spin distributions belong to the ferromagnetic phase.
The goal is to train a robust model to generalize the common properties of the domains, such as phase transition point and set of the critical exponents.

 We analysed a possible representation of the dataset used for the training/testing procedure based on traditional spin snapshots~\cite{Carrasquilla-2017} and found that it does not provide acceptable estimates of the critical temperature and correlation length exponent for interdomain transfer learning. The representation of the data as a spin correlator at a distance of half the linear lattice size $L$ proposed in~\cite{Shiina-2020} shows similar properties. 
Fortunately, the conversion of spin snapshots into energy snapshots, which are then used in the training/testing process, leads to reasonably good estimates of the critical temperature and correlation length exponent during inter-domain testing of the Baxter-Wu model using an Ising-trained neural network. Conversely, training on the Baxter-Wu dataset and testing on the Ising dataset also yields good results. Unfortunately, this does not apply to the four-state Potts model and Ising model, in which we cannot estimate either the critical temperature or the correlation length under cross-testing with energy datasets. 

In Section~\ref{sec-models} we briefly mention the Ising model, the four-state Potts model, and the Baxter-Wu model. In Section~\ref{sec-data}, we describe the data generation process in detail. In Section~\ref{sec-super}, we focus on the supervised learning and methods for analyzing model properties. In Section~\ref{sec-spin}, we present the results of our investigation of spin datasets. We discuss transfer learning in Section~\ref{sec-cross}, and in Section~\ref{sec-energy}, we present the results of our investigation of energy datasets.

\section{Statistical physics models} 
\label{sec-models}

We will focus on three two-dimensional models, the Ising model~\cite{Onsager-1944}, the Potts model~\cite{Potts}, and the  Baxter-Wu model~\cite{Baxter-1974}. The following representation of the Hamiltonians is key to the chosen representation of  dataset  energy datasets  used for training and testing in Section~\ref{sec-energy}. 

Two-dimensional Ising (IS) model on a square lattice defined by the Hamiltonian
\begin{equation}
\mathcal{H}_{IS}=-J\sum_{(i,j)}\left[\sigma_{i,j}\sigma_{i+1,j}+\sigma_{i,j}\sigma_{i,j+1}\right].
\label{eq:IS}
\end{equation}

Two dimensional 4-component Potts (4P) model on a square lattice defined by the Hamiltonian
\begin{equation}
\mathcal{H}_{4P}=-J\sum_{(i,j)} \left[\delta(\sigma_{i,j}\sigma_{i+1,j})+\delta(\sigma_{i,j}\sigma_{i,j+1})\right],
\label{eq:4P}
\end{equation}
where $\delta(\cdot,\cdot)$ is the Kroneker symbol.

Two-dimensional Baxter-Wu (BW) model on a triangular lattice defined by the Hamiltonian
\begin{equation}
\mathcal{H}_{BW}=-J\sum_{(i,j)}\left[\sigma_{i,j}\sigma_{i+1,j}\sigma_{i+1,j+1}+\sigma_{i,j}\sigma_{i,j+1}\sigma_{i+1,j+1}\right],
\label{eq:BW}
\end{equation}
and the lattice can be easily understood as a square lattice with a right-to-down diagonal. The first term in brackets is the product of the spins of the up triangle and the second term of the down triangle. 

Only ferromagnetic couplings $J{>}0$ are considered and periodic boundary conditions employed for all models. 
In the following text, we set the coupling constant $J$ equal to one, measuring the energy in units of $J$. 

The Baxter-Wu model belongs to the universality class of the four-component Potts model. The Baxter-Wu model is chosen because it lacks~\cite{BW-1973,Baxter-Book,Shchur-2010} multiplicative logarithmic corrections~\cite{Cardy-1980}, which are quite pronounced in the four-component Potts model and spoil its critical behaviour~\cite{Butera-2009}.
The Ising model belongs to a different class of universality.

Deep machine learning allows us to estimate the critical temperature $T_c$~\cite{Carrasquilla-2017} and the value of the correlation length exponent $\nu$~\cite{Chertenkov-2023}, which are known analytically and equal to 1 for the Ising model~\cite{Onsager-1944} and 2/3 for BW~\cite{Baxter-1974} and for 4P~\cite{Potts,Wu}. 

\section{Data generation}
\label{sec-data}

The data are generated with conventional Markov chain Monte Carlo methods and consist of spin configurations on the lattice~\cite{LB-book}, the so-called spin snapshots. Each spin snapshot is taken after the previous snapshot in equilibrium at temperature $T$ after a time larger than the correlation time, making them statistically independent (details can be found in the article~\cite{Chertenkov-2023}). The temperature was set in the range $T\in [T_c-0.4:T_c+0.4]$ with a step of 0.001.  We collect 1500 snapshots at each temperature for a given lattice size. The linear sizes of the lattice are $L=48,72,96,144$ and 216 with periodic boundary conditions.  The choice of these particular numbers is due to the need to preserve the symmetry of the Baxter-Wu model on finite lattices~\cite{Shchur-2010}.  

The snapshots are used to train and test NNs.  Mathematically, a spin snapshot is a $L{\times} L$ matrix, where $L$ is the linear size of the system. The elements of the matrix are the values of spins. In the case of Ising and Baxter-Wu model these are the values 1 and -1.
 
\section{Supervised learning} 
\label{sec-super}

Supervised learning is a machine learning procedure consisting of two parts: training and testing. In the specific case of our study, the neural network is trained for binary classification of snapshots for ferromagnetic and paramagnetic phases~\cite{Carrasquilla-2017}. The trained neural network is used to predict that a particular image fed to the neural network belongs to one of the two phases. In fact, the output of the neural network is provided by two neurons, the output of both of which is a number between 0 and 1. One of the neurons shows the prediction for the ferromagnetic phase, and the other for the paramagnetic phase. The sum of these predictions is equal to one, so we will only consider the prediction of belonging to the ferromagnetic phase in the future.  

\subsection{Training NN for binary classification}

We performed transfer learning between these models in all possible combinations of training and testing, using three neural network architectures: fully-connected with a single hidden layer of 100 neurons (FCNN), convolutional neural network (CNN)~\cite{CNN} and deep convolutional residual network ResNet-10 (RN)~\cite{ResNet} with 10 layers. 

 First, we trained the networks using  randomly selected $1000$ of $1500$ snapshots per temperature in each dataset for each model, and each lattice size. During training, all snapshots taken at temperature below the critical temperature $T_C$ were labeled by FM as belonging to ferromagnetic state. All snapshots taken at temperature above the critical  temperature $T_C$  were labeled by PM as belonging to the paramagnetic state. We do not provide the NN with the actual temperature of the image during training. In machine learning terms, we trained the network for binary classification of a specific model, Ising (IS), Baxter-Wu (BW), and Potts model (4P), on a specific lattice size, resulting in NN models. It is practical to refer these NN models as $FCNN_{IS,L}, FCNN_{BW,L}, CNN_{IS,L}, CNN_{BW,L}$, $RN_{IS,L}$, and $RN_{BW,L}$, where $L$  takes one of the five values mentioned in Section~\ref{sec-data}. For the Potts model for training and testing, we use three datasets that differ in their spin representation (see section~\ref{sec-spin}). Thus, the corresponding NN models are denoted, for example, as $FCNN_{4P-M,L}$, $CNN_{4P-R,L}$, etc., for a total of nine models (three datasets multiplied by three neural networks) for each lattice size $L$.

\subsection{Testing NN models}

 After training, we still have 500 unused snapshots per temperature point in each dataset. We use them to collect FM neuron predictions that a specific  snapshot $i$ taken at temperature $T$ for the IS (BW, 4P) model on the lattice with linear size $L$, which is feed-forwarded to the NN, belongs to the ferromagnetic state. These predictions of a specific NN model are denoted as  $f_i(T;L)$, $i, \; ({i=}1,2, \ldots, N)$.
The  values $f_i(T;L)$ are used to estimate corresponding probability~\cite{Carrasquilla-2017}
\begin{equation}
    P(T;L) = \frac{1}{N}\sum_{i=1}^{N}f_i(T;L)
\label{eq:PTL}
\end{equation}
and variation~\cite{Chertenkov-2023}
\begin{equation}
    V(T;L)=\frac{1}{N}\sum_{i=1}^{N}\left(f_i(T;L)\right)^2-P(T;L)^2.
 \label{eq:VTL}
\end{equation}
 In the limit of infinite $N$, the function $P(T;L)$ will be the probability that $N$ snapshots taken at temperature $T$ on the lattice with linear size $L$ belong to the ferromagnetic phase.

The analysis of these functions is used to estimate~\cite{Carrasquilla-2017}
 the critical temperature based on probabilities $P(T;L)$ and to estimate~\cite{Chertenkov-2023} the correlation length exponent based on variations $V(T;L)$.

\subsection{Critical behaviour analysis}

  We use procedure~\cite{Chertenkov-2023}, which proposed a Gaussian fit of the $V(T;L)$ function 
 \begin{equation}
    V(T;L) \propto \exp \left( -\left(T-T^*(L)\right))^2/2\sigma^2(L) \right)
  \label{eq:Gauss}
\end{equation}
and analyzes the position of the maximum $T^*(L)$ and width $\sigma(L)$ as functions of the lattice size. The position of the maximum is usually associated with pseudo-critical temperature and reflects  strong temperature driven fluctuations in the critical region of second order phase transitions~\cite{FF-1967}. It is known that the position of the maximum of thermodynamic functions, such as specific heat and magnetic susceptibility, shifts away from the critical point for finite system sizes~\cite{FF-1967,FF-1969}
\begin{equation}
 T^*(L) - T_c\propto 1/L^{1/\nu},
\label{eq:TT}
\end{equation}
where $\nu$ is the critical exponent of the correlation length, which diverges as the temperature approaches the critical value $T_c$. 
Accordingly, the width behaves as 
\begin{equation}
\sigma(L)\propto 1/L^{1/\nu}.
\label{eq:muL}
\end{equation}
Similarly, in the article~\cite{Chertenkov-2023}, it was proposed to use the  behaviour of the width ~(\ref{eq:muL})  to estimate $\nu$  and the temperature shift~(\ref{eq:TT}) to estimate the critical temperature $T_c$. Analysis of the Ising and Baxter-Wu models showed~\cite{Chertenkov-2023} that these procedures do indeed provide reasonable estimates of correlation length exponent and critical temperature for both models.
  
\section{Testing spin data sets} 
\label{sec-spin} 

In this section, we show how the procedure proposed in the article~\cite{Chertenkov-2023} and described in the previous section works for three models: IS, BW, and 4P with Hamiltonians in expressions~(\ref{eq:IS},\ref{eq:4P},\ref{eq:BW}). 
 
\subsection{Input matrix} 
 
Snapshots of spin configurations simulated for the IS and BW models are presented as $L{\times}L$ matrices with +1 and -1 entries,  denoting the Ising spins up and down.

\begin{figure}
  \center
    \includegraphics[width=\columnwidth]{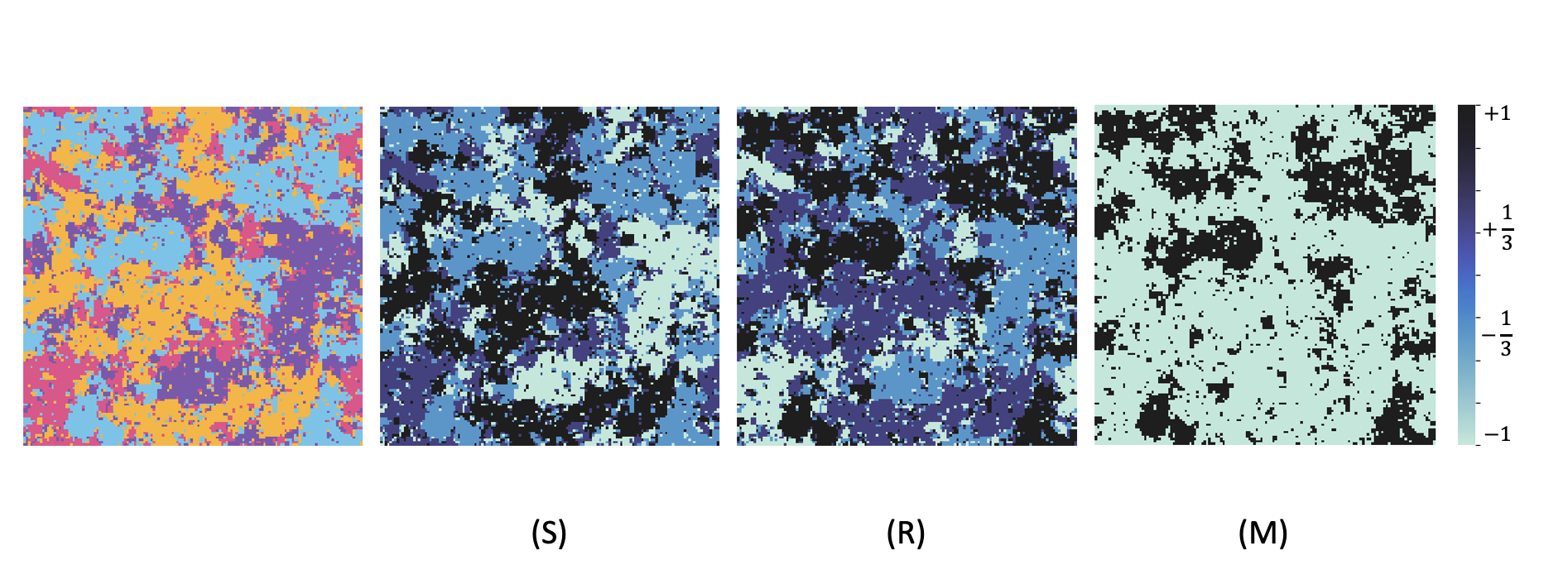}
\caption{An example of the spin configuration of the four-state Potts model. Left: the initial configuration at temperature $T=0.9107$ in the critical region, followed by simple (S), ranked (R), and majority (M) representations in the input matrix. See the text for details. }
\label{fig1}
\end{figure} 

In the case of the Potts model, there are many ways to represent the Potts spins in the input matrix. In this study, we tested three representations of spins for the 4-component Potts model:
\begin{itemize}
\item[S ] (Simple) The spin configurations $\sigma_i$ of Potts spins $\sigma_i\in\{0,1,2,3\}$ are represented in the input matrix $A$ with elements $a_i\in \{1,1/3,-1/3,-1\}$, with equidistant intervals.
\item[R] (Ranked) Count the number of spin values  $\sigma_i\in\{0,1,2,3\}$ of a particular snapshot, calculate the rank of spin $\sigma_i$ and assign the element $a_i$  of the input matrix $A$ a value from the set $\{1,1/3,-1/3,-1\}$ in descending order according to the rank of spin $\sigma_i$.
\item[M] (Majority) The elements of matrix $a_i$ are set to 1 if spin $\sigma_i$ belongs to the majority (highest rank), and to -1 otherwise.
\end{itemize}
 
Figure~\ref{fig1} shows an example of a spin configuration colored with four colors corresponding to the values of $q$ and the three possible representations in the input matrix. Simple (S) and ranked (R) representations are simply recoloring of the initial configurations, while the majority (M) representation differs from them. The black color (+1 in the color scale shown on the right)  in the M representation corresponds to the blue spins in the original configuration, which are the majority spins. The light color (-1 on the color scale) corresponds yellow, red, and purple spins in the original configuration. Note that the number of spins of the majority in representation M may be less than the number of spins of the minority -- the black area is smaller than the blue area in Figure~\ref{fig1}.

\subsection{Critical temperature estimation} 

\begin{figure}
  \center
    \includegraphics[width=\columnwidth]{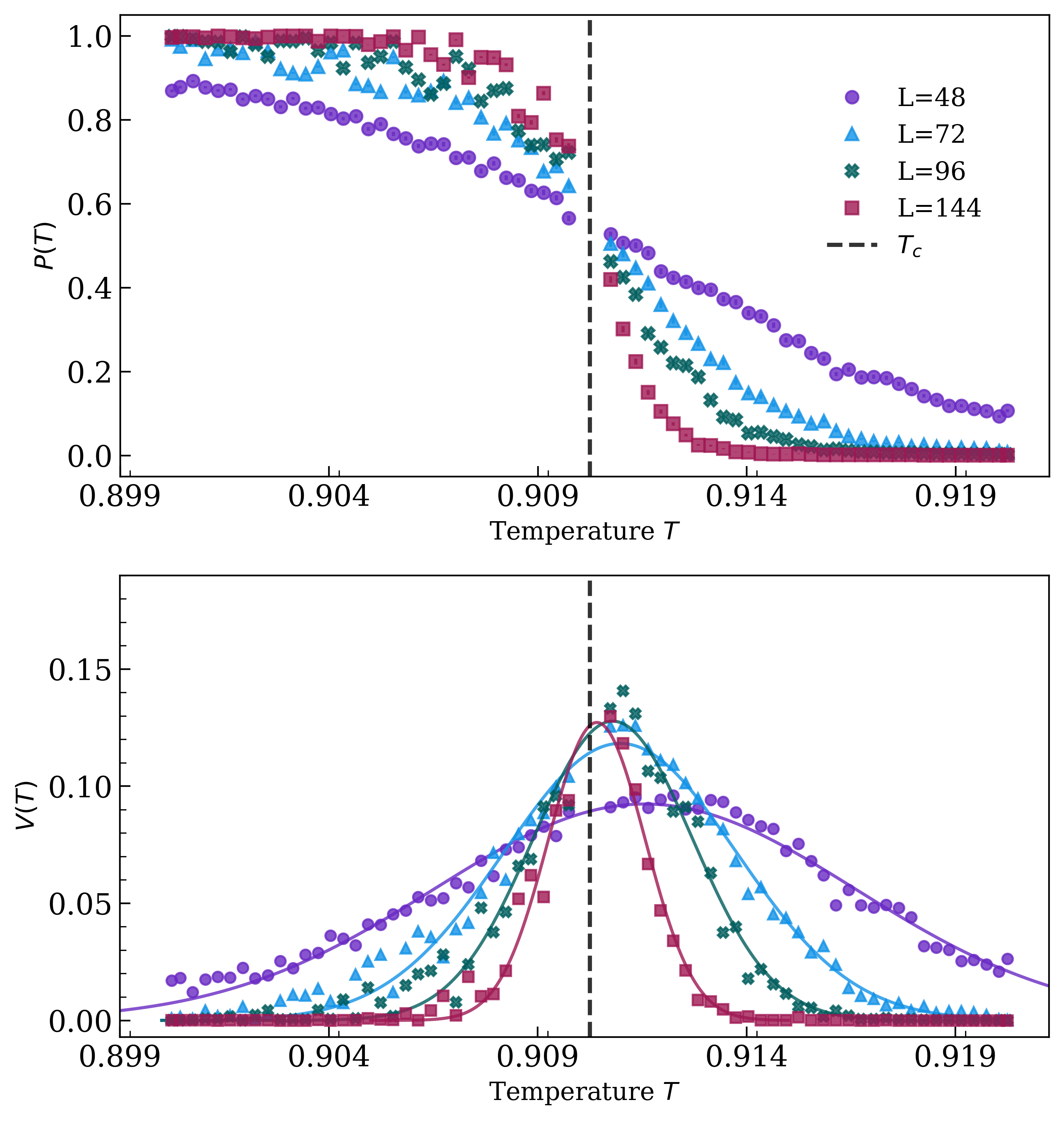}
\caption{ Probability $P(T;L)$ of the ferromagnetic phase (left) and its variation $V(T;L)$ (right), estimated for the Potts model using the dataset 4P-S and CNN.} 
\label{fig2}
\end{figure}

Figure~\ref{fig2} shows an example of calculating $P(T)$ and $V(T)$ functions using expressions~(\ref{eq:PTL}) and~(\ref{eq:VTL}) with predictions $f_i(T;L)$ as a result of testing the S-representation of 4P model configurations and using a CNN network. The position of the $V(T)$ maximum shifts toward the critical temperature in accordance with expression~(\ref{eq:TT}), which is used to estimate the critical temperature $T^*$. 
 
Tables~\ref{table1}-\ref{table3} show results of the estimation of the critical temperature of three models using the spin configurations and using NN model trained on the same spin representation in the input matrix. Critical temperature of the Ising and Baxter-Wu models coincide by chance with the value $T_c\approx 2.269185$. Critical temperature of the four-state Potts model is $T_c\approx 0.910239$. The columns labeled by $\Delta$ and $\Delta/\epsilon$ represent the difference between the estimated critical temperature $T^*$ and the exact critical temperature $\Delta{=}|T^*{-}T_c|$, divided by the statistical error $\epsilon$ of the linear fitting~\cite{footnote1}. 

Estimations of the critical temperature in all cases coincide well with the exactly known values which is not surprising due the fact that neural network trained for the binary classification on paramagnetic and ferromagnetic phases~\cite{Carrasquilla-2017}.  Note the less accurate critical temperature estimations of the 4P model using CNN and RN. What is different with the initial approach mostly used after the paper by Carrasquilla and Melko~\cite{Carrasquilla-2017} is that we estimated critical temperature from the position of the maximum of variation $V(T)$, expressions~(\ref{eq:Gauss}) and~(\ref{eq:TT}), which reflects temperature fluctuations in the critical region~\cite{Chertenkov-2023}. This gives us possibility to estimate correlation length exponent using expressions~(\ref{eq:Gauss}) and~(\ref{eq:muL}).

\begin{table}
\begin{tabular}{|l | l |  l | r || l |  l | r | } \hline
Model  &  $T^*$  &  $\Delta$ & $\Delta/\epsilon$ &  $1/\nu_\sigma$  &  $1/\nu_{\sigma^-}$ & $1/\nu_{\sigma^+}$ \\ \hline
IS &  2.2699(5) & 0.0007 &  1.4 &  1.02(1) & 1.03(13) &  0.98(3)  \\ 
BW & 2.2691(1) & 0.0001 & 1 &  1.49(2) & 1.57(2) & 1.37(8) \\
4P-S & 0.9102(5) & 0.0000 & 0 & - & - & 1.36(1) \\
4P-R & 0.9106(5) & 0.0004 & 0.7 & 1.49(4) & 1.60(9) &  1.49(4) \\
4P-M & 0.9107(8) & 0.0005 & 0.6 &  1.45(4) &  1.47(5) &  1.46(2) \\ \hline
\end{tabular}
\caption{Estimation of the critical temperature and the correlation length exponent using {\em spin configurations} by testing the Ising model (IS), the Baxter-Wu (BW) model, and four-state Potts model (4P) with a FCNN trained on the {\em same} model. See the text for the meaning of $\Delta$ and $\epsilon$. For the 4P model, the symbols S, R, and M denote the spin representation in the input matrix.}
\label{table1}
\end{table}

\begin{table}
\begin{tabular}{| l | l |  l | r || l |  l | r | } \hline
Model  &  $T^*$  &  $\Delta$ & $\Delta/\epsilon$ &  $1/\nu_\sigma$  &  $1/\nu_{\sigma^-}$ & $1/\nu_{\sigma^+}$ \\ \hline
IS &  2.2727(6) & 0.0035 &  5.8 &  1.06(3) & 1.11(5) &  1.07(2)  \\ 
BW & 2.2687(2) & 0.0005 & 2.5 & 1.44(5) & 1.54(6) & 1.48(5) \\
4P-S & 0.9102(1) & 0.0000 & 0 & 1.34(4) & - & 1.42(4) \\
4P-R & 0.9110(5) & 0.0008 & 1.5 & 1.49(6) & 1.62(3) & 1.46(6) \\
4P-M & 0.9107(8) & 0.0005 & 0.6 & 1.44(3) & 1.35(8) & 1.35(4) \\ \hline
\end{tabular}
\caption{Same as in the table~\ref{table1} but using CNN neural network architecture.}
\label{table2}
\end{table}

\begin{table}
\begin{tabular}{| l | l |  l | r || l |  l | r | } \hline
Model  &  $T^*$  &  $\Delta$ & $\Delta/\epsilon$ &  $1/\nu_\sigma$  &  $1/\nu_{\sigma^-}$ & $1/\nu_{\sigma^+}$ \\ \hline
IS &  2.2667(6) & 0.0025 &  4.1 &  1.25(3) & 1.24(7) &  1.25(3)  \\ 
BW & 2.2690(2) & 0.0002 & 1 & 1.47(5) & 1.62(13) & 1.46(4) \\
4P-S & 0.9107(5) & 0.0005 & 1 & 1.45(7)  & - & 1.17(18) \\
4P-R & 0.9101(5) & 0.0001 & 0.3 &  1.49(5) & 1.59(14) &   1.49(5)  \\
4P-M & 0.9111(3) & 0.0009 & 3.2 & 1.48(5) &  1.47(6) & 1.38(3) \\ \hline
\end{tabular}
\caption{Same as in the table~\ref{table1} but using RN neural network architecture.}
\label{table3}
\end{table}

\subsection{Correlation length exponent estimation}

 The correlation length exponent can be estimated~\cite{Chertenkov-2023} using the expression~(\ref{eq:muL}) and the results are given in Tables~\ref{table1}-\ref{table3}. The abbreviations $\nu_\sigma$,  $\nu_{\sigma^-}$, and $\nu_{\sigma^+}$ correspond to the Gaussian approximation~(\ref{eq:Gauss}) of the entire function $V(T;L)$, the left side of $V(T;L)$, and the right side of the $V(T;L)$, respectively. We did this because function $V(T;L)$ is not symmetric due to the different scales of thermal fluctuations and different values of the correlation length amplitude in the ferromagnetic and paramagnetic phases. 
 
 Estimates of the correlation length are not as stable as estimates of the critical temperature.  Recall the exact values for $1/\nu=1$ for the Ising model and $1/\nu=3/2$ for the Baxter-Wu model and the 4-state Potts model.

FCNN estimations provide a more accurate approximation of the Gaussian, especially for the entire function, as reflected in the values of $1/\nu_\sigma$ of Table~\ref{table2}. The left part of $V(T;L)$ provides less accurate estimates of the correlation length exponent. Moreover, in the case of the simple representation of the 4P model, 4P-S, it is impossible to select the Gaussian distribution for the data $V(T;L)$ due to the scattering of the data in the ferromagnetic phase, see Tables~\ref{table1}-\ref{table3}. This can be explained by the fourfold degeneracy of the ground state. Indeed, the R and M representations produce better estimation of the correlation length exponent. At the same time, estimates of the critical temperature with the M representation is less accurate. This contradicts to the widespread belief that neural network see magnetization. In our opinion, neural networks are sensitive to the correlations between spins in patterns.

\section{Transfer learning}
\label{sec-cross}

In this section, we present the results of transfer learning, which involves  testing one physical model using a neural network (NN) trained on another model. We use abbreviation BW@IS for the results of testing the Baxter-Wu model using the NN trained on the Ising model. Other abbreviations are similar. We also highlight the use of specific datasets, spin datasets, or  bond energy datasets. The neural networks used are a fully connected neural network (FCNN), a convolutional neural network (CNN), and a deep convolutional residual network (RN).

\subsection{Transfer learning with spin datasets}

We remind the reader that the neural network is trained using a spin configuration on a lattice size with linear dimension $L$. The test results allow us to estimate the probability of the ferromagnetic state $P(T;L)$ and its variation $P(T;L)$. Analysis of the finite size with an infinite lattice limit, as described in section~\ref{sec-super}, allows us to estimate $T^*$ the critical temperature and the correlation length exponent $\nu$. 

\begin{table}
\begin{tabular}{| l | l |  l | r || l |  l | r | }  \hline
IS@BW &  $T^*$  &  $\Delta$ & $\Delta/\epsilon$ &  $1/\nu_\sigma$  &  $1/\nu_{\sigma^-}$ & $1/\nu_{\sigma^+}$ \\ \hline
FCNN&  2.2756(9) & 0.0064 &  7 & -- & 1.04(3) &  1.00(1)  \\ 
CNN & 2.2817(7) & 0.0125 & 18 & -- & 0.89(9) & 0.82(10) \\
RN & 2.2714(21) & 0.0022 & 1 & -- & -- & --  \\ \hline
\end{tabular}
\caption{ Estimation of critical temperature and correlation length exponent using {\em spin configurations} by testing the Ising model (IS) with the NN model trained on the Baxter-Wu (BW) model.}
\label{table4}
\end{table}

\begin{figure}
  \center
    \includegraphics[width=\columnwidth]{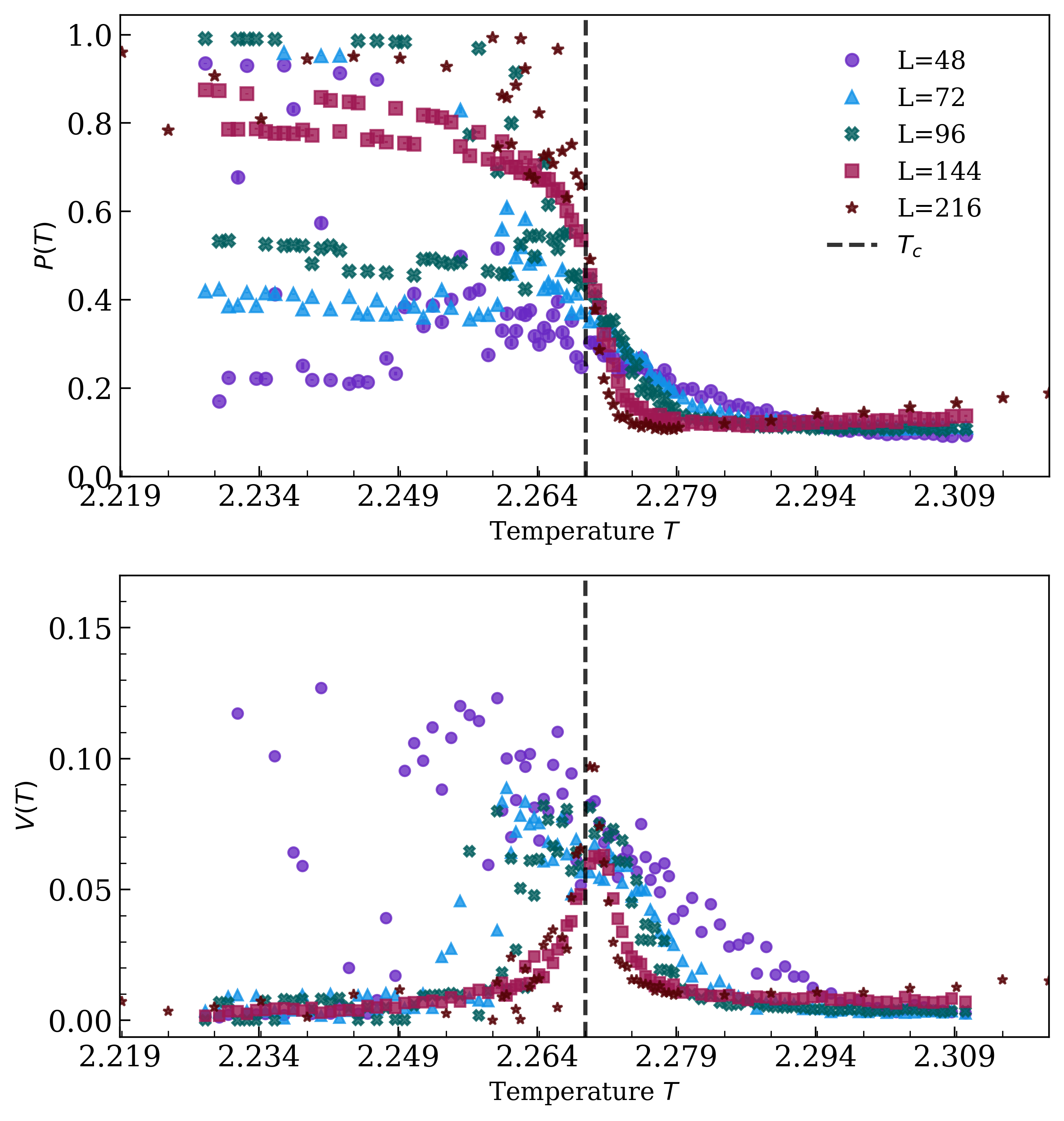}
\caption{Probability $P(T)$ and variation $V(T)$ of the ferromagnetic phase estimated for the Baxter-Wu model using NN trained with the Ising model with {\em spin dataset} and FCNN model, i.e., the transfer learning BW@IS-FCNN.}
\label{fig3}
\end{figure}

\begin{figure}
  \center
    \includegraphics[width=\columnwidth]{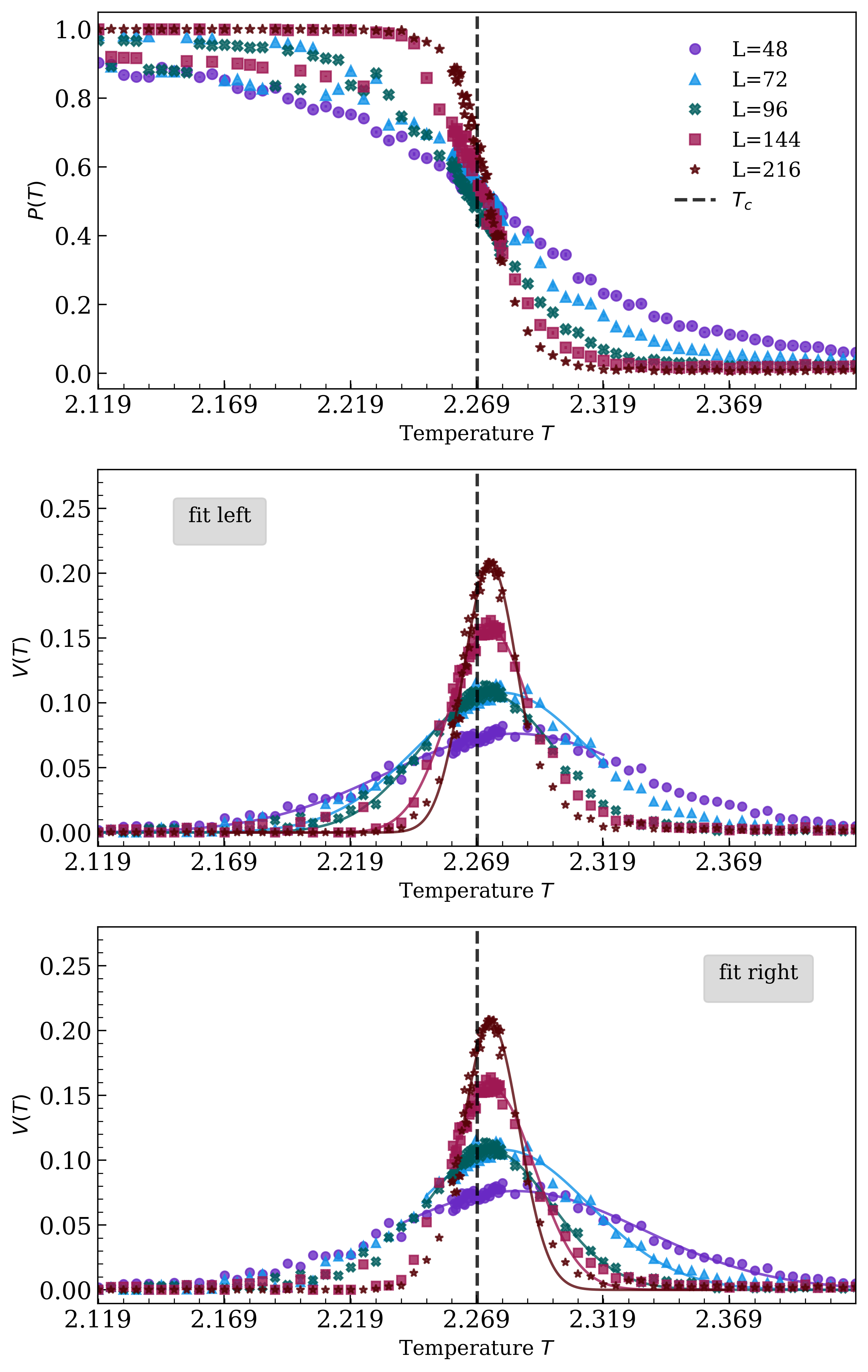}
\caption{Probability $P(T)$ and variation $V(T)$ of the ferromagnetic phase estimated for the Ising model using NN trained with the Baxter-Wu model with {\em spin dataset} and FCNN model, i.e., the transfer learning IS@BW-FCNN. Demonstration of independent fits of the left and right wings of the variation $V(T)$. }
\label{fig4}
\end{figure}

\subsubsection{Transfer learning between universality classes}

Table~\ref{table4} presents the results of the critical behavior assessment of the Baxter-Wu model using neural networks trained on the Ising model, BW@IS transfer learning case. The temperature estimation results in table~\ref{table4} can be compared with the first rows in the tables~\ref{table1}-~\ref{table3}, and they show approximately the same estimation accuracy. The correlation length estimation results are not so good, but only partially when using the FCNN network. It is not possible to extract width of the distribution $1/\nu_\sigma$ in all cases, but the estimation of the left $1/\nu_{\sigma^-}$  and right  $1/\nu_{\sigma^+}$ parts of the variation gives acceptable results with FCNN. 

The problem of knowledge transfer when using spin datasets becomes more apparent when attempting to test the Ising model using NN trained on the Baxter-Wu model, i.e., the case of IS@BW transfer learning. The Figure~\ref{fig4} shows the scattering of the probability distribution in the low-temperature phase, which is related to the difference in the symmetry of the ferromagnetic phases of BW and IS models.  

The ground state of the Baxter-Wu model is fourfold while the ground state of the Ising model is twofold. This makes temperature estimation less accurate and  correlation length estimation simply impossible. One might assume that the spin representation of the samples contains additional information about correlations and that a different representation of the configurations could improve the quality transfer learning. This turns out to be true, but not in all cases. There are other features  that can distort the  analysis, in particular logarithmic corrections to scaling. This will be demonstrated later in cross learning of four-state Potts model and Baxter-Wu model, which belong to the same universality class -- they have the same symmetry and the same critical values, but the Potts model has logarithmic corrections that are not present in the Baxter-Wu model. Therefore, the correlations differ. 

\begin{table}
\begin{tabular}{| l | l |  l | r || l |  l | r | } \hline
IS@4P-S &  $T^*$  &  $\Delta$ & $\Delta/\epsilon$ &  $1/\nu_\sigma$  &  $1/\nu_{\sigma^-}$ & $1/\nu_{\sigma^+}$ \\ \hline
FCNN&  2.2670(141) & 0.0022 & 0.2   & 1.06(4) & 1.12(5) &  1.05(1)  \\ 
CNN & 2.2580(11) & 0.0111 & 10 &  0.97(10) &  0.93(13) & 1.01(10) \\
RN & 2.2924(5) & 0.0262 & 52 & 1.08(1) & 1.26(5) &0.93(3)  \\ \hline
\end{tabular}
\caption{ Estimation of critical temperature and correlation length exponent using {\em spin configurations} by testing the Ising model (IS) with the NN model trained on the Potts model 4P-S dataset.}
\label{table5}
\end{table}

\begin{table}
\begin{tabular}{| l | l |  l | r || l |  l | r | } \hline
4P-S@IS &  $T^*$  &  $\Delta$ & $\Delta/\epsilon$ &  $1/\nu_\sigma$  &  $1/\nu_{\sigma^-}$ & $1/\nu_{\sigma^+}$ \\ \hline
FCNN&  0.9107(5) & 0.0005 & 1  & - & - &  1.20(9)  \\ 
CNN & 0.9106(5) & 0.0004 & 0.8 & - & - &0.21(6) \\ \hline\hline
4P-R@IS &  $T^*$  &  $\Delta$ & $\Delta/\epsilon$ &  $1/\nu_\sigma$  &  $1/\nu_{\sigma^-}$ & $1/\nu_{\sigma^+}$ \\ \hline
FCNN&  0.9098(5) & 0.0004 & 0.9  & -  & - &  1.06(25)  \\ 
CNN & 0.9132(8) & 0.0008 & 3.2 & - & - & 0.42(7)\\ \hline
\end{tabular}
\caption{ Estimation of critical temperature and correlation length exponent using {\em spin configurations} by testing the Potts model with two spin datasets (S and R) with the NN model trained on the Ising model dataset.}
\label{table6}
\end{table}

\begin{figure}
  \center
    \includegraphics[width=.9\columnwidth]{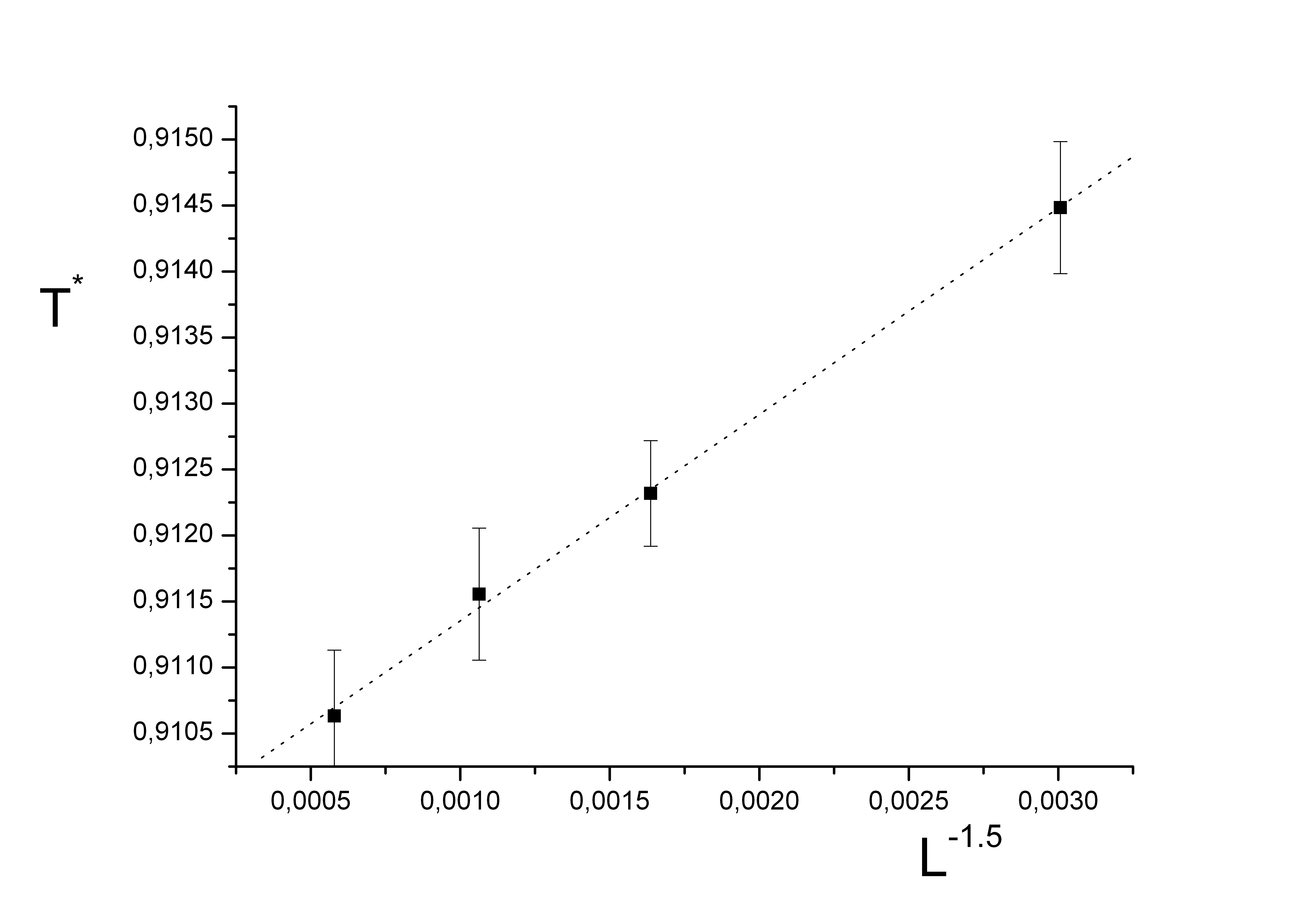}
\caption{Final scaling of the maximum position $T^*(L)$ of the function $V(T;L)$ relative to $L^{-1/\nu}$, estimated for 4P-R@IS with FCNN. The dotted line represents an approximation of the data with the results given in the first row of Table~\ref{table6}.}
\label{fig5}
\end{figure}

At the same time, the transfer learning  between four-state Potts model and Ising model using spin datasets demonstrate better results. For example, table~\ref{table5} shows critical temperature estimate and correlations length exponent estimates of the Ising model datasets using NN model trained with the Potts model 4P-S dataset, i.e., the case of IS@4P-S transfer learning. The first observation from table~\ref{table5} is that this case works better than IS@BW, although the accuracy of the temperature estimate leaves room for improvement. The second observation is that the estimates of the correlation length exponent are quite satisfactory. Thus, there is a desire to abandon the interpretation that in the case of IS@BW, only the difference in the symmetry of the ordered ferromagnetic phase can be explained. It is also possible to hypothesize that the difference lies in the difference between square and triangular lattices. We will see in the case of cross-testing on bond energies that for IS@BW there is neither the influence of the symmetry of the ground state nor the difference in the geometry of the datasets. However, in cross-testing on the Ising model with variable diagonal interaction, it was found that no deviation in the expected values is observed in a wide range of interactions~\cite{DDS-1,DDS-2} and a serious deviation is clearly visible in the case of strong anisotropy, where a rearrangement of the correlation function is known~\cite{Stephenson-1970,Wu-1976}.

\begin{figure}
  \center
    \includegraphics[width=\columnwidth]{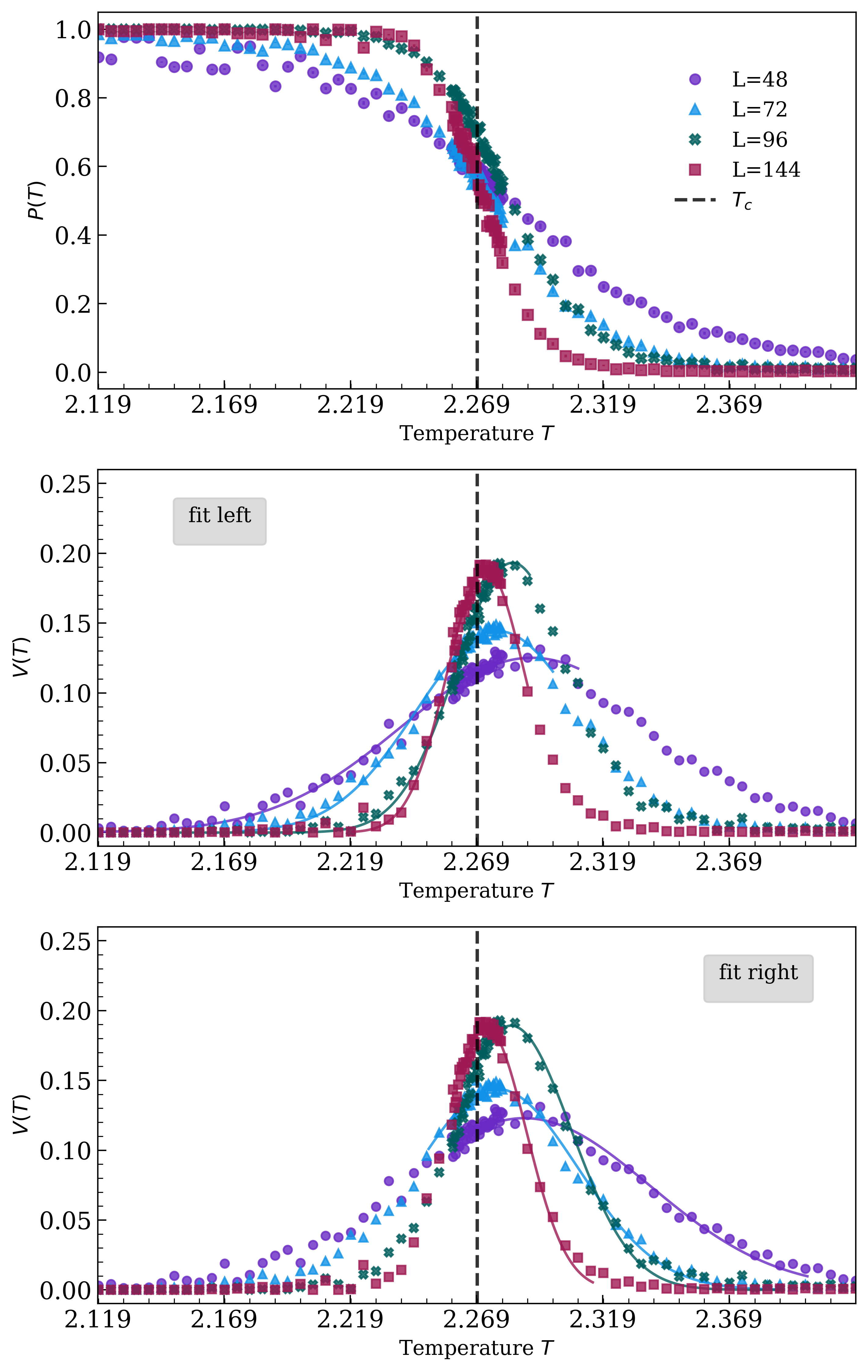}
\caption{Probability $P(T)$ and variation $V(T)$ of the ferromagnetic phase estimated for the Ising model using NN trained with Potts model S {\em spin dataset} and FCNN model, i.e., the transfer learning IS@4P-S-FCNN.}
\label{fig6}
\end{figure}

\begin{figure}
  \center
    \includegraphics[width=\columnwidth]{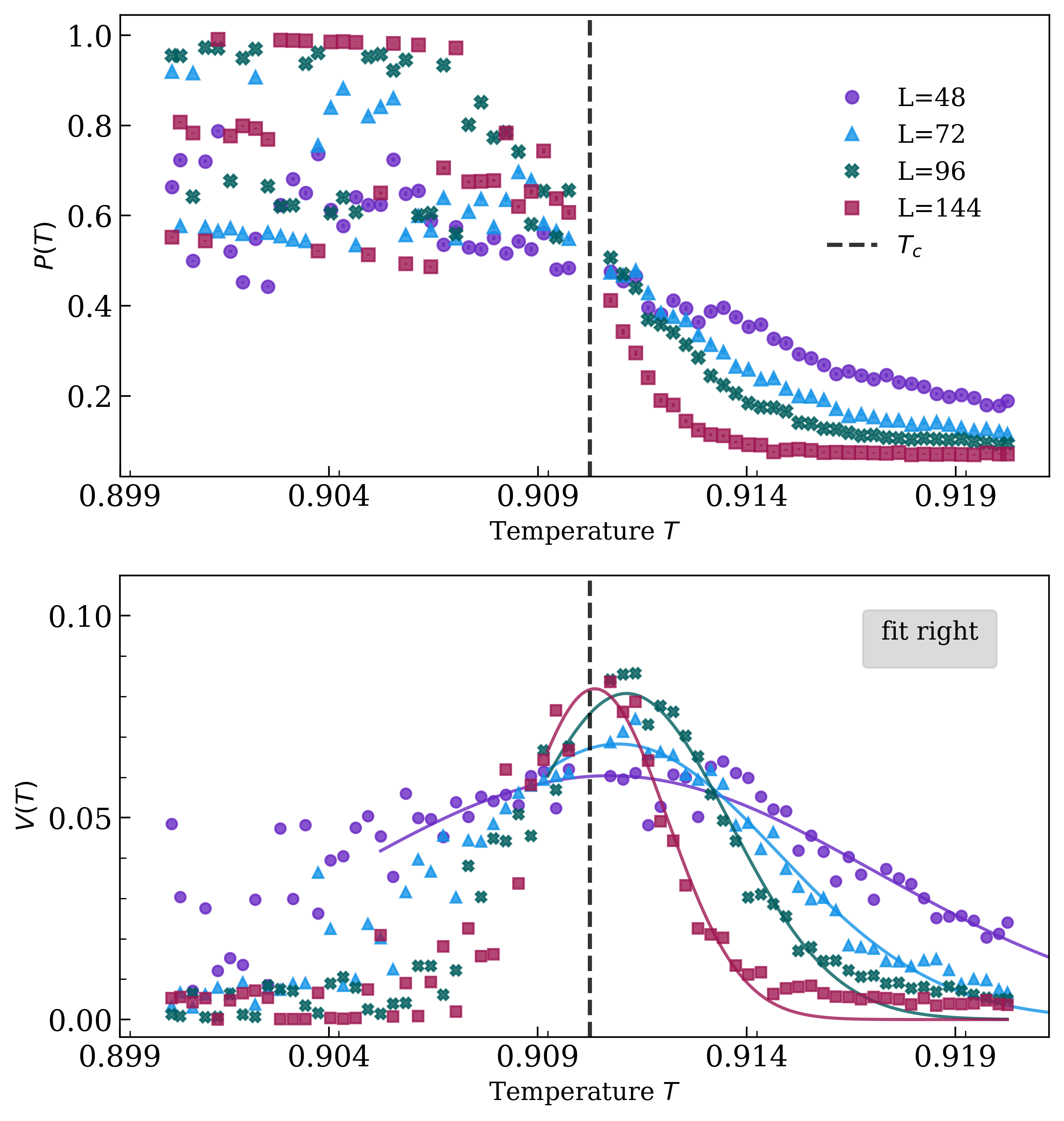}
\caption{Probability $P(T)$ and variation $V(T)$ of the ferromagnetic phase estimated for Potts model S dataset using NN trained with the Ising model {\em spin dataset} and FCNN model, i.e., the transfer learning 4P-S@IS-FCNN.}
\label{fig7}
\end{figure}

Backward learning with knowledge transfer, 4P-S@IS, and 4P-R@IS demonstrate even better results than BW@IS, at least for critical temperature estimation, as shown in table~\ref{table6} and figure~\ref{fig5}. However, the functions $P(T;L)$ and $V(T;L)$ are distributed similarly to the functions for BW@IS, so the effect of the difference in the symmetry of the ground state is still present, see Figures~\ref{fig6} an \ref{fig7}.
 
We do not present the results of testing using the RN neural network, since nothing can be extracted from the data.

\subsubsection{Transfer learning within universality class}

When transferring training of two models in the same universality class, we obtain asymmetric results. NN models trained on all three datasets (S,R, and M) of the four-state Potts model do not give reasonable results for all three networks used (FCNN, CNN, and RN) to predict the phases of the Baxter-Wu model, i.e., BW@4P did not work.

\begin{figure}
  \center
    \includegraphics[width=\columnwidth]{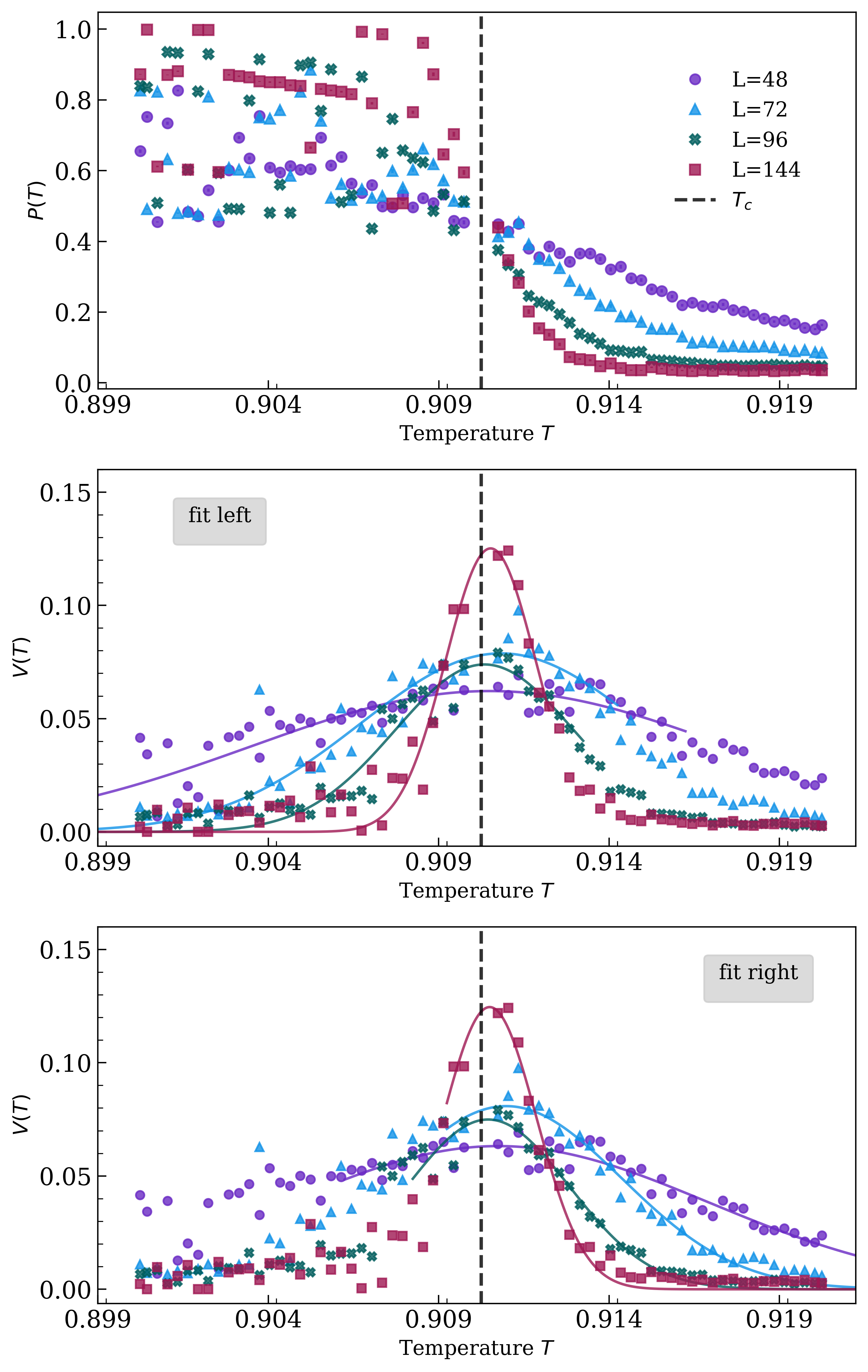}
\caption{Probability $P(T)$ and variation $V(T)$ of the ferromagnetic phase estimated for Potts model S dataset using NN trained with the Baxter-Wu model {\em spin dataset} and FCNN model, i.e., the transfer learning 4P-S@BW-FCNN.}
\label{fig8}
\end{figure}

At the same time, 4P@BW performs quite well in estimating the critical temperature, see table~\ref{table7}, the value of which differs significantly for the two models.  Figure~\ref{fig8} shows that the probability of the ferromagnetic phase $P(T;L)$ in the low-temperature phase is scattered similarly to 4P-S@IS-FCNN (see Figure~\ref{fig7}), but the variation $V(T;L)$ behaves slightly more smoothly.  Thus, we can estimate the correlation length exponent with sufficient accuracy in some cases highlighted in bold in Table 7, which are close to the exact value $1/\nu=1.5$. However, overall assessment of the correlation length exponent is not possible.

\begin{table}
\begin{tabular}{| l | l |  l | r || l |  l | r | } \hline
4P-S@BW &  $T^*$  &  $\Delta$ & $\Delta/\epsilon$ &  $1/\nu_\sigma$  &  $1/\nu_{\sigma^-}$ & $1/\nu_{\sigma^+}$ \\ \hline
FCNN&  0.9105(5) & 0.0003 & 0.6  & \bf 1.46(3) & \bf 1.52(7)  &  1.36(3)  \\ 
CNN & 0.9104(5) & 0.0002 & 0.4 & 1.22(11) & 1.26(11) & 1.2(2) \\ \hline\hline
4P-R@BW &  $T^*$  &  $\Delta$ & $\Delta/\epsilon$ &  $1/\nu_\sigma$  &  $1/\nu_{\sigma^-}$ & $1/\nu_{\sigma^+}$ \\ \hline
FCNN&  0.9116(8) & 0.0014 & 1.7   & 1.28(4) & \bf 1.5(2) &  1.23(4)  \\ 
CNN & 0.9070(5) & 0.0032 & 6.5 & 1.12(35) & 1.01(48) & \bf 1.4(4) \\ \hline
\end{tabular}
\caption{Estimation of critical temperature and correlation length exponent using {\em spin configurations} by testing the Potts model with two spin datasets (S and R) with the NN model trained on the Ising model dataset.}
\label{table7}
\end{table}

\section{Testing energy datasets}
\label{sec-energy}

\subsection{Energy snapshot datasets} 

Due to the failure of cross-domain testing using spin snapshots, we propose to use a different representation of the input datasets, the energy-based representation. The energy-based representation reflects spin interactions in the Hamiltonians~(\ref{eq:IS})-(\ref{eq:BW}).

For the Ising model datasets, using spin snapshots, we form two $L{\times} L$ matrices with elements equal to the horizontal coupling energy with elements  $e^1_{i,j}{=}{-}\sigma_{i,j}\sigma_{i+1,j}$ in the first matrix and with elements equal to the vertical coupling energy with elements $e^2_{i,j}{=}{-}\sigma_{i,j}\sigma_{i,j+1}$ in the second matrix.

For the Potts model,  the first matrix elements calculated with the first term in Expr.~(\ref{eq:4P}) $e^1_{i,j}{=}2\delta(\sigma_{i,j}\sigma_{i+1,j})-1$ and the second matrix elements calculated with the second term $e^2_{i,j}{=}2\delta(\sigma_{i,j}\sigma_{i+1,j})-1$.

For BW model datasets, the first matrix elements calculated with the first term in Expr.~(\ref{eq:BW}) $e^1_{i,j}{=}{-}\sigma_{i,j}\sigma_{i+1,j}\sigma_{i+1,j+1}$ and the second matrix elements calculated with the second term $e^2_{i,j}{=}{-}\sigma_{i,j}\sigma_{i,j+1}\sigma_{i+1,j+1}$.

These representations result in matrix elements equal to -1 or 1 for all three models. Our energy snapshot approach is an advance on the problem of domain-specific adaptation of different models, including transfer learning between universality classes.  It is found to be well suited for cross-domain learning between Ising and Baxter-Wu models.

The energy dataset is twice as large as the spin dataset. Therefore, we modified the neural networks accordingly. The CNN and RN neural networks will have a convolution input layer consisting of two input channels, while the rest of the neural networks will remain unchanged. In contrast, the FCNN neural network will have an input layer with twice as many neurons, and unfortunately, this naive approach of doubling the size of the input layer does not lead to an acceptable evaluation. Therefore, we discard the FCNN results used for analyzing the energy dataset.

\subsection{Energy datasets results} 

The results of estimates when testing all three models  with a network trained on {\em same model} and using the {\em energy datasets} is presented in the Tables~\ref{table8} and~\ref{table9} using CNN and RN neural networks. They should  be compared with the data in tables~\ref{table2} and~\ref{table3} obtained with spin datasets. The precisions are comparable within statistical errors, although we should not take the comparison too seriously, as a way of estimating the systematic error of the approach has not yet been developed.

\begin{table}
\begin{tabular}{| l | l |  l | r || l |  l | r | } \hline
Model  &  $T^*$  &  $\Delta$ & $\Delta/\epsilon$ &  $1/\nu_\sigma$  &  $1/\nu_{\sigma^-}$ & $1/\nu_{\sigma^+}$ \\ \hline
IS &  2.2660(5) & 0.0032 &  6.4  & 1.12(3) &   1.15(5) &   1.07(2)  \\ 
BW & 2.2685(4) & 0.0007 & 1.7 &  1.48(5) &   1.61(10) &   1.52(3) \\
4P& 0.9099(5) & 0.0003 & 0.7 &   1.46(2) &  1.50(2) &  1.42(4) \\ \hline
\end{tabular}
\caption{ Estimation of the critical temperature and the correlation length exponent using {\em energy dataset} by testing the Ising model (IS), the Baxter-Wu (BW) model, and four-state Potts model (4P) with a CNN trained on the {\em same model}. See the text for the meaning of $\Delta$ and $\epsilon$. }
\label{table8}
\end{table}

\begin{table}
\begin{tabular}{| l | l |  l | r || l |  l | r | } \hline
Model  &  $T^*$  &  $\Delta$ & $\Delta/\epsilon$ &  $1/\nu_\sigma$  &  $1/\nu_{\sigma^-}$ & $1/\nu_{\sigma^+}$ \\ \hline
IS &  2.2697(7) & 0.0005 &  0.7 &  1.15(3) &   1.17(6) &  1.09(11)  \\ 
BW & 2.2686(3) & 0.0006 & 2.0 &   1.50(7) &   1.62(17) &   1.52(4) \\
4P& 0.9104(5) & 0.0002 & 0.4 &   1.52(7) &  1.66(17) &   1.47(4) \\ \hline
\end{tabular}
\caption{ Same as in table~\ref{table8} but with ResNet-10 (RN) neural network. }
\label{table9}
\end{table}

After demonstrating that energy datasets can indeed be successfully used to evaluate critical properties using the supervised learning approach previously developed for spin datasets~\cite{Carrasquilla-2017,Chertenkov-2023}, we use the same transfer learning procedure described in Section~\ref{sec-cross} to cross-test the IS, BW, and 4P models with energy datasets.

The results of testing the Ising model using neural networks trained on the BW and 4P datasets are presented in Table~\ref{table10}. The results are not very encouraging and differ significantly for IS@BW and IS@4P when compared to the cross-validation results on spin models for similar cases presented in Tables~\ref{table4} and \ref{table5}. In the cases of IS@BW and IS@4P, the use of energy datasets yields better results than the use of spin datasets.  
In the case of IS@BW testing using energy datasets, CNN neural networks perform better than RN neural networks for both temperature estimation and correlation length estimation. 
At the same time, the result is better for testing IS@4P with RN than with CNN, but only for correlation length estimates and not as good for critical temperature estimates.

The opposite testing even more complicated. The BW@IS testing produces excellent estimations with energy datasets, while 4P@IS is not applicable for the correlation length exponent estimates, as can be seen from~\ref{table11}.

\begin{table}
\begin{tabular}{| l | l |  l | r || l |  l | l | } \hline
IS@BW &  $T^*$  &  $\Delta$ & $\Delta/\epsilon$ &  $1/\nu_\sigma$  &  $1/\nu_{\sigma^-}$ & $1/\nu_{\sigma^+}$ \\ \hline
CNN & 2.2714(11) & 0.0022 & 2 &  1.16(5) & 1.31(8) &  0.98(1) \\
RN &   2.2136(89) &   0.0555 & 4.4 & 0.85(6)& 1.13(11) & 0.78(15)  \\ \hline
IS@4P &  $T^*$  &  $\Delta$ & $\Delta/\epsilon$ &  $1/\nu_\sigma$  &  $1/\nu_{\sigma^-}$ & $1/\nu_{\sigma^+}$ \\ \hline
CNN & 2.6396(21) &   0.3704 &   176& 0.74(9) & 0.67(5) &   0.93(11) \\
RN &  2.4750(12) &  0.2058 &   172 &   1.03(4) &   0.97(4) &   1.03(3) \\ \hline
\end{tabular}
\caption{ Estimation of critical temperature and correlation length exponent using {\em energy dataset} by testing the Ising model (IS) with the NN model trained on the Baxter-Wu (BW) model and Potts model (4P).}
\label{table10}
\end{table}

\begin{table}
\begin{tabular}{| l | l |  l | r || l |  l | l | } \hline
BW@IS &  $T^*$  &  $\Delta$ & $\Delta/\epsilon$ &  $1/\nu_\sigma$  &  $1/\nu_{\sigma^-}$ & $1/\nu_{\sigma^+}$ \\ \hline
CNN & 2.2694(4) & 0.0002 & 0.5 &   1.45(2) &  1.51(2) &   1.42(6) \\
RN & 2.2686(4) & 0.0006 & 1.5 &  1.45(3)& \  1.45(10) &   1.47(4)  \\ \hline
4P@IS &  $T^*$  &  $\Delta$ & $\Delta/\epsilon$ &  $1/\nu_\sigma$  &  $1/\nu_{\sigma^-}$ & $1/\nu_{\sigma^+}$ \\ \hline
CNN &  0.8819(8) & 0.0283 &  35 & 0.79(4) & 0.9(4) & 0.74(5) \\
RN &  0.8912(8) & 0.0190 &  24 & 0.54(14) & 1.14(13) & - \\ \hline
\end{tabular}
\caption{ Estimation of critical temperature and correlation length exponent using {\em energy dataset} by testing Baxter-Wu (BW) model and Potts model (4P) with the NN model trained on the Ising model (IS).}
\label{table11}
\end{table}

We can highlight a partial success, as we were able to use a neural network trained on a single-class universality dataset to estimate the correlation length exponent of a model in another universality class for the first time. This was made possible by using a suitable representation of the dataset. In our case, these are datasets with bond energy configuration.

\section{Discussion}  
\label{sec-disc}

In this article, we analyzed the possibility of using a neural network trained on one physical model to test another physical model. We analyzed three models in two universality classes, the Ising model and the four-state Potts and Bxter-Wu models from another universality class.  We used three neural network architectures: FCNN, CNN, and RN, three data representations for the four-state Potts model spin configurations, and two data representations for all models—spin configurations and bond energy configurations. 

\begin{figure*}
  \center
    \includegraphics[width=2\columnwidth]{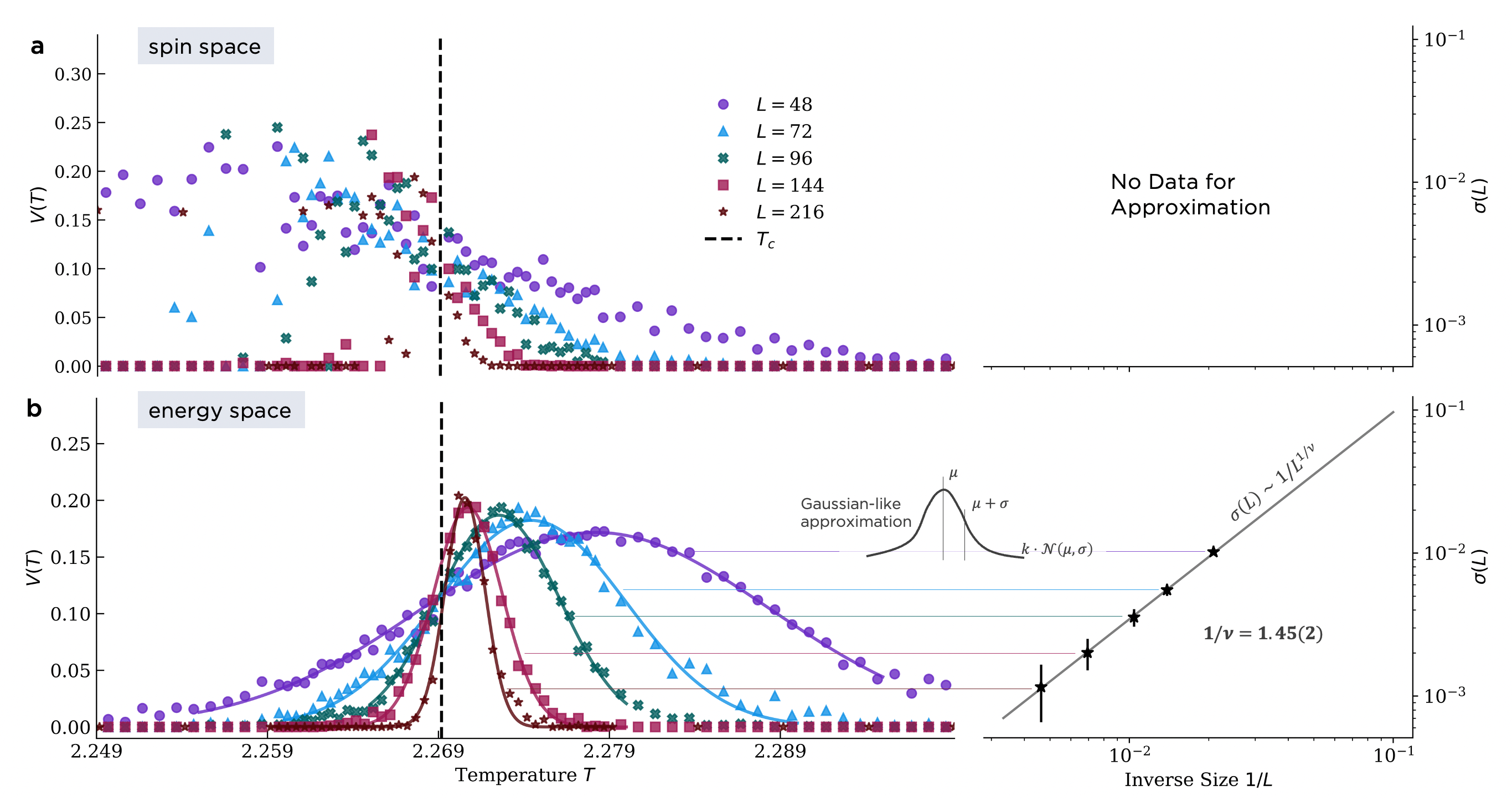}
\caption{Cross-domain transfer learning between Ising and Baxter-Wu models using CNNs. {\bf a)} The variation $V(T)$ of the FM phase prediction for the Baxter-Wu model trained on the Ising model (BW@IS) using the {\em spin domain}  cannot be approximated. {\bf b)} The variation $V(T)$ of the same testing/training combination of models BW@IS on the {\em energy domain} with Gaussian approximation and estimation of the inverse correlation length exponent $1/\nu$ from $\sigma(L)$ scaling.}
\label{fig9}
\end{figure*}

A summary of present study is provided in Table~\ref{table12} for spin dataset and in Table~\ref{table13} for energy dataset. All analyzed data representations are successfully applicable to the diagonal when testing the physical model on a neural network trained on the same model. This means that supervised learning can be successfully applied to evaluate the critical behavior of models on any type of data representation.  We also successfully performed this with the same accuracy on local spin energy and spin correlation datasets, but we do not report the results in this paper to avoid cluttering the discussion. All together, this means that neural network can ``see'' something different that only magnetization as believed by many researches. The energy datasets contain no informations on the spins and magnetization of the sample. 

The cross-testing (knowledge transfer) is not so regular as can be seen from the non-diagonal elements of Tables~\ref{table12} and~\ref{table13}. The only interesting difference is that cross-testing with energy datasets works better for the pair of cases IS@BW and BW@IS, and especially for BW@IS. 

 \begin{table}
\begin{tabular}{|l|c|c|c|} \hline
Test/Train &  Ising & Baxter-Wu & 4-Potts \\ \hline
Ising & $T_c, \nu$ & $T_c, \nu$, \bf Uns&  $T_c, \nu$ (S)  \\
Baxter-Wu & -  & $T_c, \nu$ &  -  \\
4-Potts & $T_c$ (S,R), \bf Uns  & $T_c$ (S,R,M), \bf Uns& $T_c, \nu$ (S,R,M) \\ \hline
\end{tabular}
\caption{Summary of the estimation of $T_c$ and $\nu$ while testing one physical model using NN trained on another physical model using {\em spin datasets}. {\bf Uns} stands for the unstable predictions. Abbreviations S,R, and M stands for the dataset of four-state Potts model.}
\label{table12}
\end{table}

\begin{table}
\begin{tabular}{ |l|c|c|c|} \hline
Test/Train &  Ising & Baxter-Wu & 4-Potts \\ \hline
Ising & $T_c, \nu$ & $T_c, \nu$ &  $T_c, \nu$, \bf Uns   \\
Baxter-Wu & $T_c, \nu$  & $T_c, \nu$ &  $T_c$, \bf Uns  \\
4-Potts & -   & $T_c$, \bf Uns & $T_c, \nu$  \\ \hline
\end{tabular}
\caption{ Summary of the estimation of $T_c$ and $\nu$ while testing one physical model using NN trained on another physical model using {\em energy datasets}. {\bf Uns} stands for the unstable predictions.}
\label{table13}
\end{table}

There are several possible reasons for the difficulty of cross-learning. It can be assumed that this is due to the difference in the ground state of the models, and with the help of energy datasets, we avoid this difference using energy representation for BW@IS and IS@BW.  At the same time, use of the energy dataset does not help to estimate critical exponent for BW@4P and 4P@BW when the cross-testing models in the same universality class. We have no explanation for this and can only assume that it is related to differences in lattices, square in on case and triangular in the other. However, this is not always the case, and  the cross-learning of Ising model on the square lattice with variable diagonal coupling shows that it works well in a wide range of coupling ratios and is limited only by strong antiferromagnetic couplings, where the correlation function begins to oscillate~\cite{DDS-1}.

Another possible explanation is that we are unaware of systematic errors in this approach.

The main conclusion that can be drawn from our study is that choosing an appropriate domain to represent data for the purpose of knowledge transfer in machine learning is not straightforward and obviously predictable. We had to go through many options to select a suitable domain and possible pairs of statistical physics models for the knowledge transfer process. So far, we have been able to find one pair and one data representation domain to successfully demonstrate knowledge transfer between models from different universality classes. Nevertheless, there is one such example, and it led to transfer learning between two models in two universality classes, with a satisfactory transition temperature estimate and an estimate of the critical exponent of the correlation length. This is shown graphically in Figure~\ref{fig9}.

It should be noted that many groups have previously reported critical temperature estimates using the spin representation of various models~\cite{Carrasquilla-2017,Chertenkov-2023,Alexandrou-2020,Morningstar-2018,Deng-2022,Efthymiou-2019,Hu-2017,Canabarro-2019,correlator}. This is not surprising since they used a binary classification procedure that is very sensitive to the number of training epochs, and starting from a certain number of epochs the neural network is almost perfectly trained to classify the snapshots into ordered and disordered phases, leading to a good estimate of the transition temperature. A more detailed discussion is given in~\cite{Sukhoverkhova-2023}. At the same time, the ability to extract the correlation length exponent with reasonable accuracy decreases with an increase in the number of epochs. And after approximately 5-10 epochs, this becomes simply impossible.

\section{Acknowledgments}
The first short version of this article was supported by Russian Science Foundation grant No. 22-11-00259,  and the final extended version was supported by grant No. 25-11-00158.

The simulation was carried out using the computing resources of high-performance computing at the National Research University Higher School of Economics.


\end{document}